\documentclass[aps,prl,showpacs,twocolumn]{revtex4}
\usepackage{graphicx}

\begin{document}

\title{Electronic Correlations in Oligo-acene and -thiophene Organic Molecular
Crystals}
\author{Geert Brocks$^1$, Jeroen van den Brink$^{1,2}$ and Alberto F.
Morpurgo$^3$}
\affiliation{ $^1$Faculty of Science and Technology and MESA+ Research
Institute, University of Twente, P.O. Box 217,
7500 AE Enschede, The Netherlands \\
$^2$Institute-Lorentz for Theoretical Physics, Leiden University, P.O. Box 9506,
2300 RA Leiden, The Netherlands \\
$^3$Kavli Institute of Nanoscience, Delft University of Technology, Delft, The Netherlands}
\date{\today}

\begin{abstract}
From first principles calculations we determine the Coulomb interaction between two holes on
oligo-acene and -thiophene molecules in a crystal, as a function of the oligomer length. The
electronic polarization of the molecules that surround the charged oligomer, reduces the bare
Coulomb repulsion between the holes by approximately a factor of two. The effects of relaxing
the molecular geometry in the presence of holes is found to be significantly smaller. In all
cases the effective hole-hole repulsion is much larger than the valence bandwidth, which
implies that at high doping levels the properties of these organic semiconductors are
determined by electron-electron correlations.
\end{abstract}


\maketitle

A unique aspect of organic molecular crystals is that the electronic properties
of these solids bear the marks of both molecular and condensed matter physics.
In a crystal the organic molecules preserve their identity, since molecular
crystals are held together by van der Waals interactions without the formation
of intermolecular covalent bonds. This results in a small overlap between the
wave functions of electrons located on adjacent molecules, which leads to rather
narrow electronic bands.
Because of the small electronic bandwidths in these materials,
the energy associated with interactions such as the electron-phonon or
electron-electron
interaction, may dominate over the kinetic energy of the charge carriers.

At a low density of charge carriers the interaction with phonons is expected to be dominant,
which possibly leads to the formation of polarons~\cite{Silinsh94}. At increased carrier
density --approaching one carrier per molecule-- the role of electron-electron interactions
becomes more and more important. In this regime Coulomb interactions may stabilize correlated
magnetic ground states.

Probably the most studied example of organic conductors in the high-density regime is that of
C$_{60}$. In this material it has been shown experimentally that the Coulomb repulsion
between two carriers on the same molecule is substantially larger than the electronic
bandwidth~\cite{Lof92}. At high carrier concentrations, the large Coulomb interaction results
in a strongly correlated ground state. Both K$_4$C$_{60}$ and ammoniated K$_3$C$_{60}$ are
Mott-Hubbard insulators, for instance~\cite{Knupfer97,Takenobu00}.

It can be expected that electron-electron interactions are not only important in
C$_{60}$ crystals, but also in crystals of many other molecules. In
this paper we consider oligo-acenes and oligo-thiophenes,
two classes of molecules often used in organic electronic applications,
and we study their electronic structure in the
high density regime using first principles calculations~\cite{charge}.

\begin{figure}
\includegraphics[width=6cm,keepaspectratio=true]{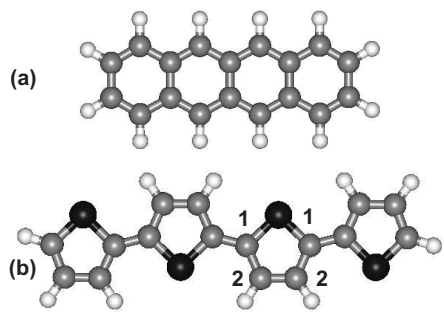}
\caption{Oligo-acenes are ladder-type hydrocarbons and oligo-thiophenes consist of a chain of
thiophene rings. The molecules (a) tetracene and (b) quater-thiophene are shown, which have
$n=4$ rings. They are planar and have (a) $D_{2h}$ and (b) $C_{2h}$ symmetry. The black, grey
and white balls represent sulphur, carbon and hydrogen atoms.} \label{fig:th4}
\end{figure}

For all the molecules of experimental interest we find that the Coulomb interaction between
two carriers on the same molecule is much larger than the electronic bandwidth, also when the
electronic screening of this interaction in the crystal is taken into account. Our results
indicate that, analogous to C$_{60}$, crystals of oligo-acenes and -thiophenes should be
viewed as correlated electron systems. In addition we find that, compared to the Coulomb
interaction, the energy associated with relaxing the molecular geometry upon charging is
smaller. This implies that the charge-charge Coulomb repulsion is the dominant factor at high
carrier density. Correlation effects in these systems should be even more pronounced than in
C$_{60}$. Many of the electronic molecular levels in C$_{60}$ are degenerate due to the high
symmetry of this molecule and such degeneracies effectively reduce the Coulomb
interaction~\cite{Gunnarsson97}. In contrast, the symmetry of oligo-acene and -thiophene
molecules (see Fig.~\ref{fig:th4}) is much lower and all their molecular levels are
non-degenerate.

\textit{Bare molecular Coulomb interactions.} We first present the results
obtained for the Coulomb interaction strength $U_{\mathrm{%
bare}}$ between holes on single, isolated oligo-acene or -thiophene molecules of different
sizes. They are indicated by $n$, the number of (benzene or thiophene) rings comprising the
oligomer, cf. Fig.~\ref{fig:th4}. To establish the value of $U_{\mathrm{bare}}$ we first
consider two singly charged $n^{+}$ ions far apart. The bare Coulomb interaction is then
determined by the process in which a single electron is moved from one molecule to the other,
i.e. $n^{+}+n^{+}\rightarrow n+n^{2+}$. Consequently the molecular Coulomb repulsion for two
holes can be expressed in terms of the total energies of the neutral, doubly and singly
charged molecules as $U_{\mathrm{bare}}=E(n)+E(n^{2+})-2E(n^{+})$.

Only ground state energies of individual molecules and ions enter here, which can be obtained
with high accuracy from first-principles electronic structure calculations. The results
presented here are obtained by density functional theory (DFT) in the local density
approximation; using a generalized gradient functional gave only small
differences~\cite{perdewzunger}. Pseudo-potentials are used to represent the atomic
cores~\cite{TM} and the valence electronic wave functions are expanded in a plane wave basis
set. Molecular calculations are performed by enclosing the molecule in a finite
box~\cite{barnett}, whereas the crystal calculations use periodic boundary conditions. This
scheme has proved to yield accurate results for molecular crystals~\cite{brocksa},
poly-thiophene~\cite{brocksb} and oligo-thiophene dimers~\cite{brocks1}.

The calculated bare Coulomb interaction $U_{\mathrm{bare}}$ between two holes on a $n$-acene
or $n$-thiophene molecule, $n$ ranging from 2 to 5 and from 4 to 16, respectively, is listed
in Table I, where the molecules are ordered according to increasing length.
$U_{\mathrm{bare}}$ ranges from $5.3$ to $1.5$ eV; it decreases as the length of the molecule
increases. On conjugated molecules the holes are delocalized over the molecule, so their
average distance is larger on a larger molecule, which of course lowers their Coulomb
repulsion.

\textit{Molecular relaxations.} So far we have not incorporated the fact that a charged
molecule relaxes its geometry, which lowers its energy. This relaxation depends on the charge
state of the molecule and effectively reduces the bare Coulomb interaction. The relaxation
energy is the energy gained when optimizing the geometry of a charged molecule, starting from
the geometry in its neutral state.

\begin{table}
\begin{tabular}{c|cccc|ccccc}
\toprule
& \multicolumn{4}{c|}{$n$-acenes} & \multicolumn{5}{c}{$n$-thiophenes} \\
\colrule
$n$ & 2 & 3 & 4 & 5 & 4 & 6 & 8 & 12 & 16 \\
\colrule $U_{\mathrm{bare}}$           & 5.26 & 4.51 & 4.01 & 3.63 & 3.54 & 2.82 & 2.46 &
1.79 & 1.48  \\
$U_{\mathrm{relax}}$          & 5.10 & 4.42 & 3.95 & 3.59 & 3.34 & 2.71 & 2.37 &
1.69 & 1.38  \\
$U_{\mathrm{relax}}^{\mathrm{AM1}}$ & 5.00 & 4.32 & 3.86 & 3.52 & 3.05 & 2.25 & 1.85 & 1.79 & 1.48 \\
$U^{\epsilon}_{\mathrm{eff}}$ & 2.47 & 1.99 & 1.75 & 1.59 & 1.47 & 1.21 & 1.09 &
0.70 & 0.56  \\
$E^{+}_{\mathrm{pol}} $     & 1.22 & 1.23 & 1.15 & 1.10 & 0.95 & 0.80 & 0.69 &
0.52 & 0.44 \\
\colrule $U_{\mathrm{eff}}$             & 2.67 & 1.96 & 1.65 & 1.39 & 1.43 & 1.10 & 0.98
& 0.64 & 0.50  \\
$W$             & 0.35 & 0.34 & 0.39 & 0.61 & 0.36 & 0.34 & 0.31 & 0.30 & 0.30
\\
\botrule
\end{tabular}
\caption{Coulomb interaction $U$ between two holes on $n$-acene and $n$- thiophene molecules.
$U_{\mathrm{bare}}$ denotes the bare interaction; $U_{\mathrm{relax}}$ and
$U_{\mathrm{relax}}^{\mathrm{AM1}}$ denote the relaxed interaction calculated with DFT and
AM1, respectively (see text, \cite{Tol}); $U^{\epsilon}_{\mathrm{eff}}$, and
$U_{\mathrm{eff}}$ denote effective interaction, calculated with the continuum model and the
discrete dipole model, respectively. $E_{\mathrm{pol}}^+$ denote the polarization energies
calculated from the discrete dipole model. The bottom row lists the valence bandwidths of the
molecular crystals. All energies are in eV.}
\end{table}

\begin{figure}
\includegraphics[width=6cm,keepaspectratio=true]{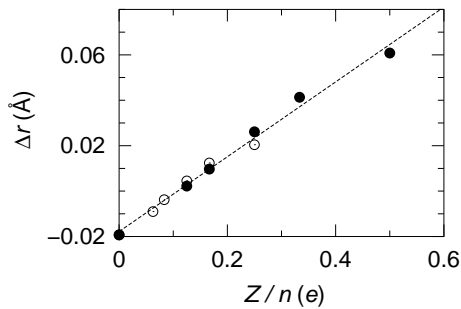}
\caption{The bond length change upon charging of $n$-thiophene molecules as function of the
average excess charge density $Z/n$. The open and closed circles represent data from singly
and doubly charged molecules; the line is a linear fit.} \label{fig:relax}
\end{figure}

As for the bare Coulomb interactions, the relaxation energies also
decrease with increasing molecular length $n$. Upon charging an
oligo-thiophene molecule, it changes from an aromatic to a quinoid
geometry~\cite{brocksa}. A good measure for characterizing the
change is to monitor the difference $\Delta r$ between the
C$_{1}$--C$_{2}$ and C$_{2}$--C$_{2}$ bond lengths, cf. Fig.
\ref{fig:th4}, where $\Delta r<0$ for the aromatic structure, and
$\Delta r>0$ for the quinoid structure. The decrease of the
relaxation energy $E_{\mathrm{relax}}$ with increasing length $n$
can be understood qualitatively assuming that the excess charge is
delocalized over the molecule. The induced change of $\Delta r$
should then roughly be proportional to the average excess charge
density, i.e. $\Delta r\propto Z/n$, where $Z=1$ ($Z=2$) for the
singly (doubly) charged molecule. Fig. \ref{fig:relax}, in which
$\Delta r$ obtained from the optimized geometry is plotted against
$Z/n$, demonstrates that the relationship is indeed approximately
linear. In a harmonic approximation the relaxation energy per
thiophene ring is proportional to $(\Delta r)^{2}$. For an
oligomer consisting of $n$ rings, we then find
$E_{\mathrm{relax}}^{Z+}(n)\propto n(\Delta r)^{2}\propto
Z^{2}/n$. The calculated relaxation energies indeed follow this
scaling. Similar geometry changes take place upon charging the
oligo-acene molecules, although the pattern is less uniform. Their
relaxation energies are much smaller, but also follow the
$E_{\mathrm{relax}}^{Z+}(n) \propto Z^{2}/n$ behavior.

Yet the absolute value of the relaxation energies is small compared to the
Coulomb interaction. The ``relaxed'' Coulomb interaction is defined as $%
U_{\mathrm{relax}}=U_{\mathrm{bare}}-(E_{\mathrm{relax}}^{2+}-2E_{\mathrm{%
relax}}^{+}).$ The inclusion of molecular relaxation upon charging changes
the Coulomb interaction by less than $10\%$; compare $U_{\mathrm{bare}}$ and $%
U_{\mathrm{relax}}$ in Table I.

Meanwhile, it has been argued that DFT calculations underestimate
relaxation energies \cite{Zuppiroli}. Therefore, we have also
calculated the latter using the AM1 semi-empirical Hartree-Fock
method \cite{Dewar}, which is known to yield a substantially
larger relaxation and is likely to give an upperbound to the
relaxation energies. The inclusion of AM1 instead of DFT
relaxation energies leads to $U_{\mathrm{relax}}^{\mathrm{AM1}}$
in Table~I. For oligo-thiophenes the AM1 relaxation effects are
indeed much larger \cite{Zuppiroli,Tol}, whereas for oligo-acenes
these effects are marginal. Note however that, irrespective of the
technique used to calculate relaxations, the Coulomb interaction
is dominant.

We therefore conclude that in the high density regime the nature of the ground state and low
energy quasi-particle excitations is foremost determined by electron-electron interaction
effects. The electron-phonon coupling is then a secondary effect which may result in a
renormalization of quasi-particle properties.

\emph{Screened Coulomb interactions.} Having discussed the response of a single molecule to
the creation of holes, we will now consider the response of the molecular crystal. In the
presence of excess charges the electronic system of the crystal reacts to effectively screen
the Coulomb interaction between the charges. The electronic screening due to electrons on the
same molecule on which the charge(s) reside is already implicitly accounted for in the
calculation of $U_{\mathrm{bare}}$. We determine the screening energy due to all molecules
that surround a charged molecule by calculating the polarization of these molecules in a
discrete dipole model, see Refs.~\onlinecite{brink1,brink2}. If we denote the total
polarization energy that results from singly/doubly charged molecules by
$E_{\mathrm{pol}}^{+/2+}$, the effective screened Coulomb interaction is given by
$U_{\mathrm{eff}}=U_{\mathrm{relax}}- (E_{\mathrm{pol}}^{2+}- 2E_{\mathrm{pol}}^{+})$.

In order to evaluate the polarization energy $E_{\mathrm{pol}}^{+/2+}$, one needs to sum over
the electric dipole moments on all molecules of the crystal lattice, induced by the excess
charge~\cite{brink1,brink2}. The strength of an induced dipole is proportional to the local
electric field, which is the sum of the field due to the charge and of that due to all the
other induced dipoles. The proportionality factor is the molecular polarizability tensor,
whose values for oligo-acenes and -thiophenes are taken from Refs. \onlinecite{munn} and
\onlinecite{andre}. The lattice sum is determined by the crystal structures, which are taken
from Refs. \onlinecite{fichou,wyckoff,holmes,mattheus}. Large molecules are not well
represented by point charges or point polarizability tensors. We therefore use the
sub-molecular approach, where the charge or polarizability is distributed over the molecule
by assigning a (sub-molecular) charge or polarizability to each benzene or thiophene
ring~\cite{Silinsh94,munn}.

One complicating factor for 4- and 6-thiophene is that at room temperature two crystal
structures exist, called the LT and HT phases~\cite{fichou}. In practice the calculated
$E_{\mathrm{pol}}^{+}$ turns out to be the same within 0.01 eV for these two phases, so we
will not list them separately. Experimental crystal structures for 12- and 16-thiophene are
not available and, for purpose of comparison only, we artificially construct their crystal
structures by starting from the HT phase of 6-thiophene and enlarging the crystal axis that
points along the long molecular axis, such as to accommodate the longer 12- or 16-thiophene
molecule.

The results for the polarization energy $E_{\mathrm{pol}}^{+}$ are shown in
Table~I. The values are between 1.2 and 0.4 eV, where increasing the length $n$
of the
molecule decreases the polarization energy. This is basically a volume effect;
for
longer molecules the charge is distributed over a larger
volume, which ``dilutes'' its electric field. Whereas the polarizability per
unit volume
of the surrounding molecules increases somewhat with the size of the conjugated
molecule,
this dependence is weak and it does not compensate for the weaker average
electric field.

The polarization energy is slightly different if one uses the ionic charge distribution given
by the AM1 calculations. This is more localized than the DFT charge distribution, leading to
a higher $E_{\mathrm{pol}}^{+}$. In practice we observe this only in the larger thiophenes,
where $E_{\mathrm{pol}}^{+}$ becomes 0.87, 0.79 and 0.72 eV for 6-, 8- and 12-thiophene.

At this point we are able to determine the effective Coulomb interaction $U_{\mathrm{eff}}$
and the results are shown in Table~I. Using AM1 charges, $U_{\mathrm{eff}}$ changes by less
than 0.06 eV even for the larger thiophenes, so we have not listed these AM1 results. One
observes that screening due to polarization of the surrounding molecules has an effect on the
Coulomb interaction which is almost an order of magnitude larger than that of molecular
relaxation. In fact, the screening is such, that the effective Coulomb interaction
$U_{\mathrm{eff}}$ is less than half $U_{\mathrm{bare}}$.

Since polarization has such a large effect, one would like to establish the sensitivity of
the results upon the model used to calculate the polarization energy. Therefore we have also
calculated the polarization energies $E_{\mathrm{pol}}^{+/2+}$ using a conceptually much
simpler, but cruder continuum model. In such a model the charged molecule is placed inside a
cavity surrounded by a homogeneous dielectric medium, where the cavity is shaped according to
the molecule. The polarization energy is then given by the electrostatic interaction between
the molecular charge distribution and the surrounding medium~\cite{brocks1}. Table I also
contains the results for the screened interaction $U_{\mathrm{eff}}^{\varepsilon }$ based
upon polarization energies calculated using a homogeneous dielectric medium. We have used a
dielectric constant $\varepsilon=3$ for the oligo-thiophenes, as well as for naphthalene
(2-acene) and $\varepsilon=3.3$ for the other oligo-acenes~\cite{munn}. One observes that the
screened interaction calculated with the crude continuum model is roughly within 10\% of the
microscopic discrete dipole model. Moreover, $U_{\mathrm{bare}}-U_{\mathrm{eff}}^{\varepsilon
}=E_{\mathrm{pol}}^{2+}-2E_{\mathrm{pol}}^{+}$ scales as $(\varepsilon -1)/\varepsilon $ so
it is not very sensitive to moderate changes of $\varepsilon $ \cite{brocks1}. We conclude
that the values of the screened Coulomb interaction obtained by the discrete dipole model are
not very sensitive to the details of that model.

\begin{figure}
\includegraphics[width=7cm,keepaspectratio=true]{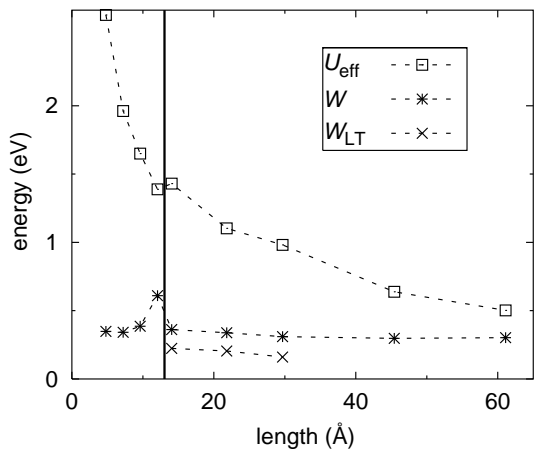}
\caption{The effective Coulomb interaction $U_{\mathrm{eff}}$ and valence bandwidth $W$
($W_{\mathrm{LT}}$ for the LT phase) as function of the oligomer length (defined as the
distance between the two extremal carbon atoms). The data for the oligo-acenes and
-thiophenes are plotted left and right of the solid vertical line.} \label{fig:u}
\end{figure}

\emph{Coulomb interaction versus bandwidth.} To determine the correlation strengths, the
effective Coulomb interaction $U_{\mathrm{eff}}$ for holes should be compared to the
bandwidth $W$ of the highest valence band. The latter can be obtained from DFT calculations
on oligo-acene and -thiophene crystals.  Crystals of the HT and LT phases of 4- and
6-thiophene have been produced, whereas 8-thiophene is only reported in the LT
phase~\cite{fichou}. Again, for purpose of comparison only, the crystal structures for the HT
phases of 8-, 12- and 16-thiophene are artificially constructed in the way discussed above.
The calculated valence bandwidths $W$ are shown in Table~I. Values in the same range have
been reported on the basis of semi-empirical calculations~\cite{Haddon95,Cornil00}.
Generally, the bandwidths depend only weakly on the size of the oligomer. For instance, it
decreases from 0.36 eV to 0.30 eV going from 4- to 16-thiophene in the HT phase. Such a weak
dependence can be explained by the observation that the highest valence band is mostly
derived from the highest occupied molecular orbital and the latter has the same character for
all oligo-thiophenes. The bandwidth in the LT phase (not shown in Table~I) decreases from
0.22 eV, 0.21 to 0.16 eV for 4-, 6- and 8-thiophene, respectively. These values are smaller
than those of the HT phase because of the different packing of the molecules in the two
phases. The bandwidths of the oligo-acenes are similar to those in the oligo-thiophenes; only
that of pentacene is substantially larger.

Fig.~\ref{fig:u} summarizes the main result of our study. The figure shows the calculated
effective Coulomb interaction $U_{\mathrm{eff}}$ and the bandwidth $W$ for the oligo- acenes
and -thiophenes, as a function of their molecular lengths. For longer molecules
$U_{\mathrm{eff}}$ becomes smaller, but does not drop below $W$ for any size of molecule. For
the molecules that are of main experimental interest at present, i.e. 4-, 6-thiophene and the
acenes, the ratio $U_{\mathrm{eff}} / W$ is between 2.3 and 7.6. So we conclude that in these
systems the intramolecular Coulomb interaction is very large compared to the electronic
bandwidth. We therefore expect that when oligo-acene and thiophene molecular crystals are
highly doped, correlation effects will be very prominent, giving rise to magnetically ordered
states.


\begin{thebibliography}{99}
\bibitem{charge} The charge on the molecules is to be compensated by counter-ions, as in alkali-doped
C$_{60}$.
\end{thebibliography}
\end{document}